\documentclass[12pt]{article}
\usepackage{graphicx}

\newcommand\pubdate{\today}

\newcommand\pubnumber

\def\Title#1{\begin{center} {\Large #1 } \end{center}}
\def\Author#1{\begin{center}{ \sc #1} \end{center}}
\def\Address#1{\begin{center}{ \it #1} \end{center}}

\newcommand\pubblock{\rightline{\begin{tabular}{l} \pubnumber\\
         \pubdate  \end{tabular}}}
\newenvironment{Abstract}{\begin{center}{\bf Abstract}\end{center} \bigskip \begin{quotation}  }{\end{quotation}}
\newenvironment{Presented}{\begin{quotation} \begin{center}
             PRESENTED AT\end{center}\bigskip
      \begin{center}\begin{large}}{\end{large}\end{center} \end{quotation}}
\def\Acknowledgements{\bigskip  \bigskip \begin{center} \begin{large}
             \bf ACKNOWLEDGEMENTS \end{large}\end{center}}

\def\beq#1{\begin{equation} \label{#1}}
\def\eeq{\end{equation}}
\def\bra#1{\left\langle #1\right\vert}
\def\ket#1{\left\vert #1\right\rangle}

\begin{document}
\begin{titlepage}
\pubblock

\vfill


\Title{{\Large\bf New Analysis of
$B \rightarrow K\pi$ data with
Pauli blocking \\
CP violation in
 $B^o$ decays, not in $B^{\pm}$ \\
SU(3) use of $B \rightarrow \pi\pi$ data invalid for $B \rightarrow K\pi$ }}
\vfill

\Author{Harry J. Lipkin}
\Address{School of Physics and Astronomy \\
Raymond and Beverly Sackler Faculty of Exact Sciences \\
Tel Aviv University, Tel Aviv, Israel\\
\vbox{\vskip 0.0truecm}
Department of Particle Physics \\
Weizmann Institute of Science, Rehovot 76100, Israel \\
and\\
High Energy Physics Division, Argonne National Laboratory \\
Argonne, IL 60439-4815, USA}


\begin{Abstract}
New data analysis with Pauli blocking explains observation of CP violation in $B^o\rightarrow K\pi$
decays, absence in $B^{\pm} \rightarrow K\pi$ decays and gives new predictions
agreeing with experiment.
Branching ratio data show pure I=1/2 amplitude predicted by pure penguin transitions
for separate relations within charged and neutral $B$ decays, but
strong violation of penguin
I=1/2 isospin relation  between charged and neutral  decays.

$B(B^o \rightarrow K^+ \pi^-) - 2B(B^o\rightarrow K^o \pi^o) =
 (0.6 \pm 1.3)\cdot 10^-6   \approx 0$

$2B(B^+ \rightarrow K^+ \pi^o) -
B(B^+ \rightarrow K^o \pi^+ ) =
(2.7 \pm 1.6)\cdot 10^-6   \approx 0$

${{\tau^o}\over{\tau^+}}\cdot 2B(B^+ \rightarrow K^+ \pi^o) -
B(B^o \rightarrow K^+ \pi^- ) =
(4.7 \pm 0.82)\cdot 10^-6 \not= 0$

This contrast between pure I=1/2
observed in individual charged and neutral final states and
I=1/2 violation in relations between them is unexpected in previous treatments.
Pauli blocking predicts this contrast by noting that two identical $u$ quarks in a relative s-wave are Pauli blocked.
Tree diagram $\bar b\rightarrow \bar s u \bar u$ for $\bar b$ producing $u$ quark at weak vertex and tree-penguin interference producing CP violation are Pauli
suppressed for $B^+$ decays with identical $u$ quark spectator. No suppression in
$B^o$ decays with spectator $d$ quark.
$B\rightarrow K\pi$ data analysis
discards all amplitudes containing two identical $u$ quarks in final state
and predicts observed isospin relations by using only Pauli-favored amplitudes ${B_u}\rightarrow
\bar s  du \bar d$ and ${B_d}\rightarrow \bar s  ud \bar u$. Pauli-favored transitions explain dependence on flavor of spectator quark which does not participate in the weak interaction. Pauli classification depends only on  final state, includes all final state interactions and all production diagrams.
Standard definition of independent color favored and suppressed tree diagrams
in  $B^\pm\rightarrow K\pi$ decays neglects Pauli blocking and misses Pauli exchange between two diagrams differing by  interchange of two identical $u$ quarks. $B^o\rightarrow K^\pm\pi^{\mp}$ diagrams have no identical quark pairs and no Pauli suppression.
Full treatment using color - spin algebra notes that Pauli blocking occurs only in $u$ quark pairs having same color and spin and confirms strong Pauli blocking in $B^{\pm} \rightarrow K\pi$ tree diagrams. No suppression in $B^{\pm} \rightarrow \pi\pi$ which also have two identical $u$ quarks but different color-spin and flavor couplings.
$B \rightarrow \pi^+\pi^o$  data useless.
Related by broken SU(3) only   to $B \rightarrow K^+P^o$ where $P^o$ is linear combination of $\pi^o, \eta, \eta'$ with only (1/4) $\pi^o$.

\end{Abstract}

\vfill

\begin{Presented}
The Ninth International Conference on\\
Flavor Physics and CP Violation\\
(FPCP 2011)\\
Maale Hachamisha, Israel,  May 23--27, 2011
\end{Presented}
\vfill

\end{titlepage}
\def\thefootnote{\fnsymbol{footnote}}
\setcounter{footnote}{0}
%


\section {Introduction - The $K-\pi$ Puzzle}
\subsection {Direct CP Violation observed in neutral $B^o\rightarrow K\pi$
decays, not in charged $B^{\pm} \rightarrow K\pi$}

A general theorem from CPT invariance shows\cite{lipCPT} that direct CP
violation can occur only via the interference between two amplitudes which have
different weak phases and different strong phases. This  holds also for all
contributions from new physics beyond the standard model which conserve CPT.

Direct CP violation has been experimentally observed\cite{PDG,HFAG}
in $B_d \rightarrow K^+ \pi^-$ decays.
\beq{acp0}
A_{CP}(B_d \rightarrow K^+ \pi^-)= -0.098 \pm 0.013
\eeq
 The experimental observation (\ref{acp0}) and the knowledge that the penguin
amplitude is dominant for the decay\cite{Ali} require that the decay amplitude must
contain at least one additional amplitude with both weak and strong phases
different from those of the penguin.

The CP violation (\ref{acp0}) has been attributed to the interference between the large contribution from the dominant penguin diagram and smaller contributions from tree diagrams.  The failure to observe CP violation in charged
decays\cite{Ali} has been considered to be a puzzle\cite{nurosgro,ROSGRO}
because changing the flavor of a spectator quark which does not participate
in the weak decay vertex is not expected to  make a difference.
 \beq{acp+}
\begin{array}{ccl}
\displaystyle
A_{CP}(B^+ \rightarrow K^o_S \pi^+)= 0.009 \pm 0.029
\hfill\\
\\
A_{CP}(B^+ \rightarrow K^+ \pi^o)= 0.051 \pm 0.025
\end{array}
\end{equation}
\subsection {New experiments sharpen The $K-\pi$ Puzzle}
New experimental results show that
the individual branching ratios for  $B^o $ and $B^+$ decays agree with
the pure I=1/2 amplitude predicted by a penguin diagram.
\beq{newpuz}
\begin{array}{ccl}
\displaystyle
B(B^o \rightarrow K^+ \pi^-) - 2B(B^o\rightarrow K^o \pi^o) =
 (19.4 \pm 0.6)- 2\cdot (9.4 \pm 0.6) = 0.6 \pm 1.3  \approx 0
\hfill\\
2B(B^+ \rightarrow K^+ \pi^o) -
B(B^+ \rightarrow K^o \pi^+ ) = (25.8\pm 1.2) - (23.1 \pm 1.0)
=2.7 \pm 1.6  \approx 0
\end{array}
\eeq
where $B$ denotes the branching ratio in units of $10^-6$.

However, the isospin relation between $B^+ $ and $B^o$ decays predicted by the penguin diagram
is in strong disagreement with experiment.
\beq{newpuz2}
{{\tau^o}\over{\tau^+}}\cdot 2B(B^+ \rightarrow K^+ \pi^o) -
B(B^o \rightarrow K^+ \pi^- ) =
4.7 \pm 0.82
\eeq
where $\tau^o/\tau^+$ denotes the ratio of the $B^o$ and $B^+$ lifetimes and we have used
the experimental values
\beq{newtestex}
\begin{array}{ccl}
\displaystyle
B(B^o \rightarrow K^+ \pi^-)  =
19.4 \pm 0.6
 \hfill\\
\\
\displaystyle
{{\tau^o}\over{\tau^+}}\cdot B(B^+ \rightarrow K^o \pi^+) =
{{(23.1 \pm 1.0)}\over{1.07}}=
21.6\pm 0.93
\hfill\\
\\
\displaystyle
{{\tau^o}\over{\tau^+}}\cdot 2B(B^+ \rightarrow K^+ \pi^o)
= {{2\cdot (12.9\pm 0.6)}\over{1.07}}=
24.1\pm 0.56
\hfill\\
\\
\displaystyle
B(B^o\rightarrow K^o \pi^o) =
(9.4 \pm 0.6)
\end{array}
\end{equation}

The relation (\ref{newpuz2}) shows  that the $B\rightarrow K\pi$ transition is not a pure penguin.  The relation (\ref{newpuz}) shows  that the $I=1/2$ prediction by a pure  penguin diagram is violated only in the ratio of the branching ratios for charged and neutral decays. The branching ratios for the individual charged and neutral decays satisfy $I=1/2$

The significant difference between the experimental values of expressions (\ref{newpuz}) and (\ref{newpuz2})
is not expected in the conventional analyzes. The two relating branching ratios for individual charged and neutral decays still vanish here while one relating charged and neutral case is finite. This indicates a surprising cancelation and motivates a search for a theoretical explanation.

\section {Puzzle resolved by calculation including Pauli principle}
\subsection {SU(3) breaking prevents using parameters from $B\rightarrow \pi\pi$
decays for $B\rightarrow K\pi$}
Standard treatments \cite{nurosgro,ROSGRO} of charmless B decays have used data from
$B\rightarrow \pi\pi$ decays together with SU(3) flavor symmetry to obtain parameters
for analysis of $B\rightarrow K\pi$. At that time precise $B\rightarrow K\pi$ data were not yet
available. New more precise data revealed contradictions with this approach\cite{nuhuor1}.
The source of these contradictions can be seen as due to SU(3) breaking.

 A final $\pi^o\pi^+$ state $\ket{f;\pi^o\pi^+}$ is a pure $I=2$ state in a pure SU(3)
 27-dimensional representation of flavor SU(3).

In the symmetry limit the strange analog of the $\pi^o\pi^+$ state in a pure SU(3)
27 is $K^+ V_{10}$ state where $V_{10}$ denotes the $V$ spin analog of the $\pi^o$
with $V=1,V_z=0 $. This state is badly broken by SU(3) symmetry breaking into
$K^+ \pi^o$, $K^+ \eta$ and $K^+ \eta'$. The $K^+ \pi^o$ state has only (1/4) of the
SU(3) 27 related to the $B\rightarrow \pi\pi$ decay. The remaining (3/4) is classified in
other representations of SU(3) which are not related to the $B\rightarrow \pi\pi$ decay.
Thus there is no possibility for using SU(3) with $B\rightarrow \pi\pi$ decay to
obtain parameters for analysis of $B\rightarrow K\pi$.

\subsection {Detailed symmetry and Pauli analysis}
The dependence on spectator flavor arises from the Pauli blocking
by the spectator quark of a quark of the same flavor participating in the weak vertex. The u-quark produced by a
tree diagram is Pauli blocked by the spectator $u$ quark in $B^+$ decay but
is not affected by the spectator $d$ quark in neutral decays. This difference in Pauli
blocking suppresses the tree contribution and CP violation in charged $B$ decays but allows
tree-penguin interference and enables CP violation to be observed in neutral decays.

The decay of a $\bar b$ antiquark to a charmless final state is described by the
vertex
\beq{Bvert}
\bar b \rightarrow
\bar q n \bar n
\end{equation}
where $n \bar n$ denotes a nonstrange $u \bar u$ or $d \bar d$ quark-antiquark pair
and $\bar q$ denotes a  $\bar d$ antiquark for $\pi\pi$ decays or a
$\bar  s$ antiquark for $K\pi$ decays.

Symmetry restrictions from the Pauli principle arise when a nonstrange spectator
quark has the same flavor as the $n$ quark emitted from the weak vertex.
In the tree diagram for $B^+$ decays both the nonstrange $n$ quark
and the spectator quark are $u$ quarks.
\beq{tree}
B^+ =\bar b u \rightarrow
\bar q u \bar u u
\end{equation}

A flavor-symmetric $uu$ state in a spatially symmetric S-wave is required by the Pauli
principle to be
antisymmetric in color and spin.
The  antiquark  pair in (\ref{Bvert}) must also be antisymmetric in color and spin.
Although no Pauli principle forbids a symmetric color - spin state
such states cannot combine with the $uu$ pair to make the spin-zero color singlet
final state $\pi\pi$ or $K\pi$.
The
fragmentation of a $uu\bar u \bar q$ state into a
$\pi^+\pi^o$ or $K^+\pi^o$ is a strong interaction which conserves flavor SU(3) and
charge conjugation.

Both the $uu$ diquark and the $\bar u \bar q$ antidiquark are thus
antisymmetric in color and spin. The generalized Pauli principle requires each
to be symmetric in flavor SU(3) and its  SU(2) subgroup isospin for $\pi\pi$ decays or
V-spin for $K\pi$ decays. Each is therefore respectively in
the symmetric isospin state with $I=1$ or in the symmetric V-spin state with $V=1$.
A final state must be even under  general charge conjugation to decay into two
pseudoscalar mesons in an orbital S wave. Thus
the $(I=1,I_z=+1)$ diquark and the $(I=1,I_z=0)$ antidiquark must be coupled
symmetrically to  $(I=2,I_z=+1)$.
Similarly the $(V=1,V_z=+1)$ diquark and the $(V=1,V_z=0)$ antidiquark must be coupled
symmetrically to  $(V=2,V_z=+1)$.
These states are in the 27-dimensional representation of flavor SU(3).

The final states in the 27 are produced from a a $u$ quark pair in a color-spin state which
satisfies the Pauli Prinicple. Final states of two pseudoscalar mesons in other representations
of $SU(3)_{flavor}$ are Pauli suppressed.

A final $\pi^o\pi^+$ state $\ket{f;\pi^o\pi^+}$ is a pure $I=2$ state in a pure SU(3)
27.
Thus the tree diagram for the nonstrange transition
$(B^+ \rightarrow \pi+ \pi^o)$ is not Pauli suppressed.

A final $K^o\pi^+$ state $\ket{f;\pi^oK^+}$
has no $V=2$ component, since both the $K^o$ and $\pi^+$
have V=1/2. Thus the tree diagram for the $K^o\pi^+$ decay must vanish and this
decay is pure penguin.

The final $K^+\pi^o$ state contains a $\pi^o$ which is a linear combination of
$V=0$ and $V=1$ states with probability of 1/4 for $V=1$.
The component with $V=0$ cannot combine with a $V=1$ $K^+$ to make $V=2$.
The $V=1$ component can   combine with a $V=1$ to make $V=2$  with a
probability of 1/2.
Thus the probability that the final
$K^+\pi^o$ state has a $V=2$ component is 1/8.
Thus we see that Pauli blocking suppresses the tree diagram for the
$(B^+ \rightarrow K^+ \pi^o)$
transition by a factor 8.

Present data are consistent with complete suppression but evidence for a
partial suppression is still down in the noise.

The $ud\bar u \bar s$ state created in the tree diagram for
$B_d$ decay has no such restrictions. It can be in a flavor SU(3)
octet as well as a 27. Its ``diquark-antidiquark" configuration includes the
flavor-SU(3) octet constructed from the spin-zero color-antitriplet
flavor-antitriplet ``good" diquark found in the  $\Lambda$ baryon and its
conjugate ``good" antidiquark. These ``good diquarks" do not exist in the
corresponding $uu\bar u \bar s$ configuration.

We again see that the Pauli effects produce a drastic dependence on spectator
quark flavor in the tree diagrams for $B \rightarrow K \pi$ decays.

Tree-penguin interference can explain both the presence
of CP violation in neutral decays and its absence charged decays.

\subsection{An approximate quantitative treatment of Pauli effects in $B \rightarrow K \pi$ decays.}

The transition from an initial $B$ meson state  consisting of a $\bar b$ antiquark and a nonstrange spectator quark
to a strange charmless two-meson final state is written
\beq{secquanx}
\begin{array}{ccl}
\displaystyle
\ket{B_d}=\bar b d \rightarrow
\bar s  \cdot \left[d \bar d+ u\bar u+\xi \cdot u\bar u \right] d
= \bar s  \cdot \left[\kappa d \bar d+ (1+\xi)\cdot  ud \bar u \right] d\approx
\bar s  \cdot (1+\xi)\cdot  ud \bar u
\hfill\\
\\
\ket{B_u}=\bar b u \rightarrow
\bar s  \cdot \left[d \bar d+ u\bar u+\xi \cdot u\bar u \right] u
=\bar s  \cdot \left[d \bar d+ \kappa\cdot (1+\xi ) u\bar u \right] u\approx
\bar s  \cdot du \bar d
\end{array}
\end{equation}
where the final state is first written as the sum of an isoscalar
$q\bar q$ pair and a $u\bar u$ pair together with a strange
antiquark and a spectator quark. This is analogous to the conventional
description as the sum of a penguin contribution and a tree contribution. The parameter $\xi$ generally considered to be small expresses the ratio of the tree and penguin contributions. The parameter $\kappa$ is a Pauli factor expressing the probability that the two nonstrange quarks are not in the same color-spin state.

Our  full color-spin analysis gives
$\kappa \approx \frac{1}{8}\approx 0$.
In this approximation a $u$ quark produced by a weak interaction cannot enter the same state as a
$u$ spectator quark.
The states with $\kappa=0$ have no quark pairs of the same flavor. We call these Pauli-favored states.
The approximate equality at the RHS of eqs.(\ref{secquanx}) sets $\kappa=0$.

The CP violation is proportional to $\xi$.
In some approximation we can write
\beq{secquanxa}
\xi \approx \frac{V_{bu}}{V_{bc}}
\end{equation}
We then see that when $\kappa=0$ the final state in neutral decays depends upon $\xi$ while final state in charged decays is independent of $\xi$. This solves one puzzle by suppressing the tree contribution and CP
violation in  charged B decay while allowing it in neutral decays.
This suppression is lost in conventional treatments which consider color-favored and color-suppressed
tree amplitudes as independent without considering Pauli suppression.

We now note that the $ud$ pair in the final states must be isoscalar by the generalized Pauli principle. The final states must then be pure isospin eigenstates with $I=1/2$. In the standard treatments\cite{nurosgro,ROSGRO} the $I=3/2$ component is not suppressed, except in pure penguin transitions

\subsection{The difference between charged and neutral decays}
We now show explicitly how the Pauli principle can forbid the tree-penguin interference and CP
violation in  charged B decays and allow them in neutral  decays.

The transition between an initial $B$ meson state into a $K \pi$ final state is described as a weak transition (\ref{secquanx}) followed by a fragmentation process in which the two $q \bar q$ pairs are combined to produce the $K \pi$ final state. The transition matrix element $\bra{K\pi}T\ket{\bar b Q}$ for the decay of a B meson can be written
\beq{frag}
\bra{K\pi}T\ket{\bar b Q }=
\bra{K\pi}F \ket{\bar s
  \cdot \left[d \bar d+ u\bar u+\xi \cdot u\bar u \right] \cdot Q}\bra{\bar s  \cdot \left[d \bar d+ u\bar u+\xi   \cdot u\bar u \right]\cdot Q}W\ket{\bar b Q }
\end{equation}
where $Q$ denotes a nonsrange quark, either $u$ or $d$, $\ket{\bar b Q}$ denotes a $b$ meson state with constituents
$\bar b$ and $Q$,
$F$ and $W$ denote respectively the fragmentation and weak transition operators. Substituting the two spectator flavors $u$ and $d$ then gives
\beq{frag2}
\begin{array}{ccl}
\displaystyle
\bra{K\pi}F \ket{\bar s
  \cdot \left[d \bar d+ u\bar u+\xi \cdot u\bar u \right] \cdot u}= \bra{K\pi}F \ket{\bar s
  \cdot \left[d \bar d+ \kappa u\bar u+\kappa u\xi \cdot u\bar u \right] \cdot u}
\hfill\\
\bra{K\pi}F \ket{\bar s
  \cdot \left[d \bar d+ u\bar u+\xi \cdot u\bar u \right] \cdot d}= \bra{K\pi}F \ket{\bar s
  \cdot \left[\kappa d \bar d+ u\bar u+ u\xi \cdot u\bar u \right] \cdot d}
\end{array}
\end{equation}
The relative phase of the two terms is left open and does not affect
the final result. Then
\beq{secquanx2}
\begin{array}{ccl}
\displaystyle
B^+ \rightarrow
\bar s  \cdot \left[d \bar d + \kappa \cdot (1+\xi) u \bar u \right] u=
\bar s  \cdot \left[d \bar du + \kappa \cdot (1+\xi) u u \bar u \right]  =K^o \pi^+ \pm \frac {1-\kappa \cdot (1+\xi) }{\sqrt 2} \cdot K^+ \pi^o
\hfill\\
B^o \rightarrow
\bar s  \cdot \left[\kappa \cdot d \bar d + (1+\xi) u \bar u \right] d=
\bar s  \cdot \left[\kappa \cdot d \bar dd + (1+\xi) ud \bar u \right]  =
(1+\xi)\cdot K^+ \pi^- \pm \frac {1+\xi-\kappa}{\sqrt 2} \cdot K^o  \pi^o
\end{array}
\end{equation}
As a simple approximation we  set $\kappa = 0$  and obtain

\beq{secquanx3}
\begin{array}{ccl}
\displaystyle
{B^+ \rightarrow
\bar s  \cdot \left[d \bar d + (1+\xi) u \bar u \right] u=
\bar s \bar d \cdot d  u}=\left[K^o \pi^+ \pm \frac {K^+ \pi^o}{\sqrt 2} \right]
\hfill\\
{B^o \rightarrow
\bar s  \cdot \left[d \bar d + (1+\xi) u \bar u \right] d=
\ bar s  \bar u\cdot (1+\xi) u d}=
(1+\xi)\cdot \left[K^+ \pi^- \pm \frac {K^o  \pi^o}{\sqrt 2}  \right]
\end{array}
\end{equation}
The transitions for the neutral decays are seen to depend upon the parameter $\xi$ while the charged transitions are independent of  $\xi$. The parameter $\xi$ is proportional to the strength of the tree amplitude. The tree amplitude produces a $u \bar u$ pair in the $b$ decay vertex. This amplitude is Pauli suppressed in charged $B$ decays where the spectator quark is also a $u$ quark. Thus in the $\kappa = 0$ approximation tree-penguin interference which might produce CP violation is present in neutral decays and absent in charged decays. This can explain how CP violation can be drastically changed by changing the spectator quark
and the otherwise mysterious result  (\ref{acp+}).
\subsection {The Sum and Difference Rules}
The standard treatment\cite{approxlip,approxgr,Gronau,ketaprimfix} assumes that four $B\rightarrow K\pi$ branching ratios
are determined by three parameters, the dominant penguin diagram $P$  and two interference terms $P\cdot T$ and $P\cdot S$ between the dominant penguin and the color-favored and color suppressed tree
diagrams. This treatment assumes the two tree contributions shown in Figs. 2 and 3 are independent and neglects Pauli blocking. It also assumes that the two tree amplitudes are sufficiently small to be
treated in first order. Second order terms $T\cdot T$, $S\cdot T$
and $S\cdot S$ are assumed to be negligible. These assumptions lead to a sum rule.
\beq{sumruleapp}
R_L \equiv 2{{\Gamma(B^+ \rightarrow K^+ \pi^o) + \Gamma(B^o
\rightarrow K^o \pi^o)} \over {\Gamma(B^+ \rightarrow K^o \pi^+ )
+  \Gamma(B^o
\rightarrow K^+ \pi^-)}} \approx 1
\eeq

The agreement\cite{Ali} with experiment\cite{PDG,HFAG} confirms these assumptions  \cite{approxlip,approxgr,Gronau,ketaprimfix}.

We now investigate what is observable in the experimental data, how to separate
the signal from the noise, how to find the`other amplitude and examine what can we learn
about it from experiment.
The sum rule (\ref{sumruleapp}) has been rearranged \cite{ketaprimfix}
to obtain a ``difference rule"
\beq{eqapp}
{{\tau^o}\over{\tau^+}}\cdot \left[ 2B(B^+ \rightarrow K^+ \pi^o)
- B(B^+ \rightarrow K^o \pi^+ )\right] \approx
B(B^o \rightarrow K^+ \pi^-)  - 2B(B^o\rightarrow K^o \pi^o)
\eeq
where  the result was expressed in terms of  branching ratios, denote
by B().

This relation (\ref{eqapp}) states that the $I=3/2$ contributions to charged and neutral
decays are equal.
Combining this result (\ref{eqapp}) with the approximate experimental result (\ref{newpuz})
confirms the result
 with smaller experimental errors and shows that in this approximation the $I=3/2$ contributions to
both charged and neutral decays vanish.

\section {Comparison with other approaches}

Previous analyses \cite{nurosgro,ROSGRO} were performed at a time when experimental
values for $B\rightarrow K\pi$ branching ratios were not
sufficiently precise to enable a significant test of the sum rule
(\ref{sumruleapp}). Values of each of the three interference terms
were statistically consistent with zero.
The full analysis required the use of data from $B\rightarrow \pi\pi$ decays and
the assumption of $SU(3)_{flavor}$ symmetry. Contributions of the electromagnetic
penguin diagram were included and the relevant CKM matrix elements were included. But
there was no inclusion of constraints from the Pauli principle nor contributions from
final state interactions.

The present analysis uses new experimental data which enable a statistically significant
evaluation of the interference terms (\ref{newpuz})
without additional information
from $B\rightarrow \pi\pi$ decays or the assumption of $SU(3)_{flavor}$ symmetry.
Contributions from all isospin invariant finite state
interactions are included as well as constraints from the Pauli principle. A
flavor topology definition \cite{nuhuor1} of the interference parameters includes contributions from the electromagnetic penguin diagram since the  quark states in final state of a photon can be rewritten as the sum of an isoscalar and
a $u\bar u$ state. However the flavor topology parameters are no longer simply related to the
CKM matrix elements. Additional assumptions and information are necessary to determine the CKM matrix elements
and explain CP violation.

The main advantage of this approach is that it gives simple explanations for the  absence of CP violation (\ref{acp+}) in charged B decays, the observed absence of an $I=3/2$ component in the final state,
the experimental value
 (\ref{newpuz}) and the vanishing of the experimental value
\beq{vanish}%
\frac{2B(B^o\rightarrow K^o \pi^o) - B(B^o \rightarrow K^+ \pi^-)} {{{[\tau^o}/{\tau^+]}}\cdot[ B(B^+ \rightarrow K^o \pi^+) + 2B(B^+ \rightarrow K^+ \pi^o)] -
2B(B^o \rightarrow K^+ \pi^- )}
=0.09\pm 0.1
\eeq

This vanishing of tree-penguin interference $B^+$ decays is explained by a symmetry analysis including the constraints of the Pauli principle on states containing a pair of identical $u$ quarks.
\section {Conclusion}

Experiment has shown that the penguin-tree interference contribution in $B^+
\rightarrow K^+\pi^o$ decay is very small and may even vanish. The
corresponding interference contributions to neutral $B\rightarrow K\pi$ decays
have been shown experimentally to be finite. In charged decays the
previously neglected Pauli antisymmetrization  produces a cancelation
between color-favored and color-suppressed tree diagrams which differ by the
exchange of identical $u$ quarks. This explains the smallness of penguin-tree interference and
small CP violation in charged $B$ decays. Pauli cancelation does not occur in neutral decay diagrams
which have no pair of identical quarks.
 This can explain why CP violation
has been observed in neutral $B \rightarrow K\pi$ decays and not in charged
decays
\section*{Acknowledgements}

This research was supported in part by the U.S. Department of Energy, Division
of High Energy Physics, Contract DE-AC02-06CH11357. It is a pleasure to thank
Michael Gronau, Yuval Grossman, Marek Karliner, Zoltan Ligeti, Yosef Nir,
Jonathan Rosner, J.G. Smith, Frank Wuerthwein and Guohuai Zhu
for discussions and
comments.

%
\catcode`\@=11 
\def\references{
\ifpreprintsty \vskip 10ex
%
\hbox to\hsize{\hss \large \refname \hss }\else
\vskip 24pt \hrule width\hsize \relax \vskip 1.6cm \fi \list
{\@biblabel {\arabic {enumiv}}}
{\labelwidth \WidestRefLabelThusFar \labelsep 4pt \leftmargin \labelwidth
\advance \leftmargin \labelsep \ifdim \baselinestretch pt>1 pt
\parsep 4pt\relax \else \parsep 0pt\relax \fi \itemsep \parsep \usecounter
{enumiv}\let \p@enumiv \@empty \def \theenumiv {\arabic {enumiv}}}
\let \newblock \relax \sloppy
 \clubpenalty 4000\widowpenalty 4000 \sfcode `\.=1000\relax \ifpreprintsty
\else \small \fi}
\catcode`\@=12 

\end{document}